\documentclass{article}
\usepackage{amsmath,amssymb}
\usepackage{type1cm}

\newcommand{\be}{\begin{equation}}
\newcommand{\ee}{\end{equation}}
\newcommand{\f}{\frac}
\newcommand{\s}{\sqrt}
\newcommand{\p}{\partial}
\newcommand{\lm}{\mathcal{L}}

\newcommand{\bea}{\begin{eqnarray}}
\newcommand{\eea}{\end{eqnarray}}
\newcommand{\ba}{\begin{align}}
\newcommand{\ea}{\end{align}}
\newcommand{\taub}{\overline{\tau}}

\newcommand{\Lc}{{\cal L}}
\newcommand{\zb}{\overline{z}}

\begin{document}

\begin{titlepage}
\thispagestyle{empty}
\begin{flushright}
IPMU13-0023

YITP-13-4
\end{flushright}

\vspace{.4cm}
\begin{center}
\noindent{\Large \textbf{An Entropy Formula for Higher Spin Black Holes via Conical Singularities}}\\
\vspace{2cm}
Per Kraus$^{a,}$\footnote{e-mail: pkraus@ucla.edu} and Tomonori Ugajin $^{b,c,}$\footnote{e-mail: tomonori.ugajin@ipmu.jp}

\vspace{1cm}
  {\it   $^{a}$ Department of Physics and Astronomy, \\
University of California, Los Angeles, CA 90095,USA,\\
 $^{b}$Kavli Institute for the Physics and Mathematics of the Universe,\\
University of Tokyo, Kashiwa, Chiba 277-8582, Japan\\
$^{c}$Yukawa Institute for Theoretical Physics,
Kyoto University, \\
Kitashirakawa Oiwakecho, Sakyo-ku, Kyoto 606-8502, Japan,\\
\vspace{0.2cm}
 }
\vskip 2em
\end{center}

\vspace{.5cm}

\begin{abstract}
We consider  the entropy of higher spin black holes  in 2+1 dimensions using the
conical singularity approach. By introducing
a conical singularity along a non contractible cycle and carefully evaluating its contribution to the Chern Simons action,
we derive a simple expression for the entropy of a general stationary higher spin black hole.
The resulting formula is shown to satisfy the first law of thermodynamics, and  yields agreement with previous results based on integrating the first law.

\end{abstract}

\end{titlepage}
\newpage
\tableofcontents
\section{Introduction}

Higher spin gravity theories  in anti de Sitter space \cite{Vasiliev:1999ba,Bekaert:2005vh}  have played an important role in
exploring new versions of holography, e.g. \cite{Klebanov:2002ja,Sezgin:2003pt,Giombi:2009wh,Giombi:2010vg,
Gaberdiel:2010pz,Campoleoni:2010zq,Henneaux:2010xg,Gaberdiel:2011wb,Gaberdiel:2012uj}. The theories themselves are interesting because
they are expected to be toy models of string theory in large curvature spacetimes.
In 2+1 dimensions, higher spin fields are topological,  as is the gravitational field, and so we can formulate the theory  in terms of
Chern Simons theory \cite{Witten:1988hc,Achucarro:1987vz,Blencowe:1988gj}.

Recently, black holes with higher spin charges in 2+1 dimensional higher spin theories
have been extensively studied \cite{Gutperle:2011kf,Ammon:2011nk,Castro:2011fm,Tan:2011tj,David:2012iu,Kraus:2012uf,Kraus:2011ds,Banados:2012ue,Campoleoni:2012hp,
Chen:2012pc,Chen:2012ba,Perez:2012cf,Tan:2012xi,Gaberdiel:2012yb,Perez:2013xi}, and it turned out \cite{Ammon:2011nk} that they do not have event horizons in general since the metric in higher spin theories
is gauge dependent. Nevertheless, there does exist a gauge invariant notion of the regularity of the (Euclidean) configuration, and one can  assign
thermodynamic properties  to the black holes \cite{Gutperle:2011kf}.
In \cite{Gutperle:2011kf} the black hole entropy is derived  by demanding it  satisfy the first law of thermodynamics.  Via the AdS/CFT correspondence, these black holes map to generalized thermal ensembles of CFTs with higher spin symmetries, and where comparison is possible the black hole and CFT entropies are found to agree \cite{Kraus:2011ds,Gaberdiel:2012yb}.  

Despite this progress, some aspects of  higher spin black holes are still unclear. First of all, a general
formula which calculates the entropy of all higher spin black holes is not known. It is important to find an analog  of the  Wald formula \cite{Wald:1993nt} for the higher spin theory.
Second, there are in fact a few different approaches to deriving the entropy of higher spin black holes, and the assumptions and results do not always agree \cite{Gutperle:2011kf,Banados:2012ue,Campoleoni:2012hp,Perez:2013xi}.   What is the relation between these different approaches?
Finally, a gauge invariant understanding of the causal structure of higher spin black holes is unavailable in general \cite{Ammon:2011nk,Kraus:2012uf}.

One approach to black hole entropy in ordinary gravity is the conical deficit method  \cite{Banados:1993qp,Iyer:1995kg,Solodukhin:2011gn,Fursaev:1995ef}.  Recall that the entropy is obtained from the partition function as
\ba
S(\beta) = -\left(\beta {\partial \over \partial \beta}-1\right) \log Z(\beta)~.
\label{Sform}
\end{align}
Applied to black holes there are two possible  of this formula.  In general, the partition function is obtained from the action of the Euclidean black hole with time periodicity $\beta$.  In the first interpretation one considers a smooth Euclidean metric at arbitrary $\beta$, evaluates the action, and then differentiates as above.  To respect smoothness, the parameters of the black hole such as the mass must vary along with $\beta$.    In the second interpretation, one keeps all parameters fixed while varying $\beta$.  Changing $\beta$ away from its preferred value thereby introduces a conical singularity at the horizon.   Carefully evaluating the contribution of this singularity to the action and evaluating (\ref{Sform}) one obtains an alternative expression for the entropy.   It turns out that these two approaches yield the same answer for an arbitrary diffeomorphism theory of gravity, and in particular they coincide with the Wald entropy \cite{Banados:1993qp,Iyer:1995kg,Solodukhin:2011gn,Fursaev:1995ef}.

One advantage of the conical singularity approach is that it makes it manifest that the entropy is associated with the local geometry at the horizon.   Another interesting aspect is that it is closely linked to methods used in computing entanglement entropy.
 When we compute the entanglement entropy  $S_{A}$ of region A in the time slice, we first introduce a deficit angle $\delta=2\pi (1-n)$ on $\p A$ which is the boundary of the region A. Then we evaluate the partition function $Z_{n}$ of the theory on the singular manifold with the deficit angle. Entanglement entropy is derived by taking the derivative of $Z_{n}$ \cite{Nishioka:2009un},
  \be
 S_{A}=\left(n\f{\p}{\p n}-1 \right)\Big |_{n=1} \log Z_{n}
\ee
Within the context of AdS/CFT the connection is even sharper, as eternal black holes can be regarded as pure states in a tensor product of two CFTs, and the entropy arises after tracing over one copy \cite{Israel:1976ur,Balasubramanian:1998de,Maldacena:2001kr}.


The concept of the regularity  of a configuration in the 3d higher spin theories is correctly defined as triviality of the holonomy of the connection \cite{Gutperle:2011kf,Castro:2011iw}. This is a direct generalization
of the regularity of the metric in the spin 2 case. Actually, at least for  spin 3 black holes, the holonomy condition and the usual regularity condition are equivalent  in the gauge
where the metric of the black hole has  an event horizon \cite{Ammon:2011nk}.

Then we expect that we can derive a Wald like formula by evaluating the action of the higher spin black hole with a conical singularity.
In this paper we generalize the conical singularity method to the  Chern Simons gauge theory in 2+1 dimensions. A sketch of the procedure is as follows: starting from
a black hole connection with trivial holonomy around the timelike cycle which satisfies the correct thermodynamical relation, we
deform the period of the Euclidean time direction so that the holonomy around the cycle becomes slightly nontrivial. The non triviality of the holonomy indicates that the field strength of the corresponding connection has a delta function divergence along a non contractible cycle. We evaluate the Chern Simons action of the singular connection by appropriately regularizing the connection, following the approach of \cite{Fursaev:1995ef}  in the metric formulation. We find the following  general formula for the entropy of a stationary higher spin black hole,
\be
S=-2\pi i k   {\rm Tr} \left[ A_{+}\left( \tau A_{+} - \bar{\tau}A_{-} \right) \right] -2\pi i k  {\rm Tr} \left[ \overline{A}_{-}\left( \tau \overline{A}_{+} - \bar{\tau}\overline{A}_{-} \right) \right] .
\label{result}
\ee
Here $A$ and $\overline{A}$ denote the two Chern-Simons gauge fields.  The spacetime coordinates are $x^\pm$ and the radial coordinate $\rho$.   The formula for $S$ is independent of $\rho$.
As far as we are aware, this is the first paper which  generalizes the conical singularity approach to the connection formalism of gravity.

To verify that this formula is physically sensible we show that it obeys the following first law variation
\ba
\delta S& = -4\pi^2 i \sum_{s=2}^\infty \alpha_s  \delta W_s  -4\pi^2 i \sum_{s=2}^\infty \overline{\alpha}_s  \delta \overline{W}_s ~,
\label{delS}
\end{align}
where $W_s$ denote the spin-s charges, and $\alpha_s$ their conjugate thermodynamic potentials (and the same for the barred versions).    Previously \cite{Gutperle:2011kf}, the black hole entropy was derived by assuming (\ref{delS}), and then integrating.   Here we find that the solution to this problem is given by (\ref{result}).   It follows immediately that (\ref{result}) will reproduce previous results based on integrating the first law.

This note is organized as follows. In section \ref{sec:mreview} we review the method to evaluate the Einstein-Hilbert action of a metric with conical singularity, and apply it to
 derive the entropy formula of black holes.  In section \ref{subsecCS} we adapt the procedure to the connection formalism.  In section \ref{firstlaw} we demonstrate that the resulting entropy formula satisfies the correct first law variation for an arbitrary higher spin black hole.   This confirms  that the  entropy formula derived in this way correctly reproduces all previous results based on integrating this first law. 
  Appendix A contains some computations related to the general stationary black hole.   

\vspace{.2cm}
\noindent
{\bf Note:}  As this manuscript was being prepared the paper \cite{deBoer:2013gz} appeared, which  arrives at our entropy formula by a different route  and shows that it obeys the first law.  This paper also greatly clarifies the relation between the different approaches to computing the entropy of higher spin black holes.

\section{Review of the derivation of the Wald formula via conical singularity} \label{sec:mreview}

In this section, we review the derivation of the Wald formula by evaluating the gravitational action
of a conical singularity. In subsection \ref {subsec:2dcon}, we evaluate the Einstein-Hilbert action of a conical singularity in two dimensions by taking the limit of a regularized metric. Since a Euclidean black hole with deficit angle on the bifurcation surface looks like a direct product of
a two dimensional cone and the bifurcation surface near the tip, we can use the result of  \ref {subsec:2dcon}
to  evaluate the action of the singular black hole. We then discuss its relation to the Wald formula in subsection \ref {subsec:rwald}.

\subsection{Evaluation of the action of a conical singularity in 2 dimension} \label{subsec:2dcon}
In this section, we review the evaluation of the
Einstein-Hilbert action of a metric with a conical singularity on a two dimensional manifold \cite{Banados:1993qp,Iyer:1995kg,Solodukhin:2011gn,Fursaev:1995ef}.
As an example, consider the metric:
\be
ds^2=e^{\Phi(r)} (dr^2+r^2d\theta ^2). \label{eq;sinm}
\ee

 Let us assume  $ \theta \sim \theta + 2\pi \alpha,~ \alpha \neq 1$. The metric has a conical singularity at the tip when the period of the $\theta$ direction is not $2\pi$.
It is convenient to embed the cone into $\mathbb{R}^3$ by the map:
\be
x=r\alpha \sin \f{\theta}{\alpha} \qquad y=r\alpha \cos \f{\theta}{\alpha} \qquad z= \s{1-\alpha^2} r \label{eq:map},
\ee
so the cone is mapped to the surface
\be
\f{1-\alpha^2}{\alpha^2}(x^2+y^2)-z^2=0
\ee
with metric $ds^2=e^{\Phi(r)}(dx^2+ dy^2 +dz^2)$, the pull back of which gives the metric (\ref{eq;sinm}) on the 2d plane. In this form, the singularity is manifest because $\f{\p z}{\p x}, \f{\p z}{\p y}$ are indeterminate
at the tip of the cone $z=0$.

Since the metric (\ref{eq;sinm}) has a conical singularity, the curvature of the metric contains a delta functional divergence at the tip in addition to the ordinary regular part,
$R \sim R_{reg}+(1-\alpha) \delta (r)$. The coefficient in front of the delta function  is attached so that  for a regular $\alpha=1$ metric, only $R_{reg}$ appears in $R$.

To see this, we would like to evaluate the contribution of the singularity to the  Einstein Hilbert action  $ \int \!\s{g}R$.
Two steps are needed. First we construct a family of smooth metrics $g_{\mu \nu}(a)$, each labeled  by a real positive number $a$, and demand
 that they  approach to the original singular one in the limit $a \rightarrow 0 $.
We call the family the ``regularization" of the singular metric. One way to construct a regularization is by modifying the surface (\ref{eq:map}) by
a function $f(r,a)$ as
\be
z=\s{1-\alpha^2}f(r,a),\qquad \p_{r} f(r,a)|_{r=0}=0, \qquad \lim_{r \rightarrow \infty} f(r,a) \rightarrow r,
\ee
with the ambient metric  $ds^2=e^{\Phi(r)}(dx^2+ dy^2 +dz^2)$  held fixed. We also assume  $f(r,0)=r$. For example, if we take
\be
z=\s{a^2+\f{1-\alpha^2}{\alpha^2}r^2},
\ee
then the surface is replaced by a smooth hyperboloid when $a \neq 0$. For general  $f(r,a)$, the pull back of the metric on the 2D plane  is
modified as
\be
ds^2=e^{\Phi(r)}(u(r)dr^2+r^2d\theta ^2) \qquad u(r)=\alpha^2 +(1-\alpha^2) (\p_{r} f(r,a))^2. \label{eq:re}
\ee
We can see these metrics are  regular at the tip $r=0$.
Second, we evaluate the Einstein Hilbert actions $I(a)$
of the regularized metrics for general non zero $a$. Since these metrics are regular, we can safely evaluate $I(a)$,
\begin{align}
I(a)&=2\pi \alpha \int^{\infty}_{0} dr \f{u'(r)}{u^{\f{3}{2}}}-\int^{\infty}_{0} dr\int^{2\pi \alpha}_{0} r d\theta \s{u(r)}\Delta \Phi(r) \nonumber \\
&=4 \pi (1- \alpha)-\int^{\infty}_{0} dr\int^{2\pi \alpha}_{0}r d\theta \s{u(r)}\Delta \Phi(r) \label{eq:conc},
\end{align}
where $\Delta$ is the Laplacian of the metric without conformal factor (depending on $u(r)$).
The  $a \rightarrow 0$ limit correspond to the action of the original metric with a conical singularity.
\be
\lim_{a \rightarrow 0}I(a)=4 \pi (1- \alpha)-\int^{\infty}_{0} dr\int^{2\pi \alpha}_{0}r d\theta  \Delta \Phi(r) \label{eq:lim}.
\ee
Since the second term of the expression can be derived by directly substituting the metric (\ref {eq;sinm}) into the action, it describes the contribution from the regular part of the curvature to the action. However, there is an additional term. The first term in (\ref {eq:lim}) is interpreted as the contribution of the conical singularity at the tip, because
it vanishes when $\alpha=1$. The result show that the scalar curvature of the metric (\ref {eq;sinm}) is given by
\be
\sqrt{g} R= \sqrt{g} R_{reg}+\f{2 (1-\alpha)}{\alpha} \delta(r),
\ee
as we expected.
It is important to note that the
contribution of the singularity to the action does not depend on the regularization function $f(r,a)$ we use. This assures us that the value $4\pi  (1- \alpha) $ is intrinsic to the conical singularity.

\subsection{Relation to black hole entropy}  \label {subsec:rwald}
We can  generalize the result to higher dimensions if the manifold we consider is the direct product of a 2 dimensional cone $C_{\alpha}=S^1_{\alpha} \times \mathbb{R}$
and a smooth manifold  $\Sigma$. In the case  of the metric of the form
\be
ds^2=e^{\Phi(r)} (r^2 d\theta^2+dr^2)+ ds_{\Sigma}^2,
\ee
we are assuming $ \theta \sim \theta + 2\pi \alpha,~ \alpha \neq 1$. Demanding that the volume of the $S^1_{\alpha} \times \Sigma$ located at $r=r_{0}$ is held fixed, the general regularization of the metric
can be written
\be
ds^2=e^{\Phi(r)} (r^2 d\theta^2+u(r)dr^2)+ ds_{\Sigma}^2, \quad u(0)= \alpha^2, \quad u(\infty)=1 \label{eq:mmet}
\ee
Note that only  $g_{rr}$ is allowed to change.
This turn out to be the correct regularization and from this
one can compute various geometric invariants in the presence of the conical deficit. The results turn out to be equivalent
to the statement that the Riemann tensor  contains a delta functional singularity on $\Sigma$,
\be
R^{\mu \nu}_{\alpha \beta}=( R_{reg})^{\mu \nu}_{\alpha \beta}+2\pi (1-\alpha)(n^{\mu}_{\alpha} n^{\nu}_{\beta}-n^{\mu}_{\beta}n^{\nu}_{\alpha}) \delta _{\Sigma}, \label{eq:rs}
\ee
where $\delta _{\Sigma}$ is delta function on $\Sigma$ which satisfie
\be
\int_{C_{\alpha} \times \Sigma} \delta  _{\Sigma}  \sqrt{g} = \int_{\Sigma}\sqrt{h},
\ee
and $n^{\mu}_{\alpha}=n^{\mu}_{1} n_{\alpha 1}+n_{2}^{\mu }n_{\alpha 2}$, where $n_{1},n_{2}$ denotes the  vector fields normal to $\Sigma$ .

 Euclidean black holes with temperature different from the Hawking temperature  are examples
of these geometries. In this case $\Sigma$ is the bifurcation surface of the black hole.

Suppose $Z(\beta )$ is the quantum gravity partition function   with fixed temperature $\beta $,
evaluated  semiclassically as
\be
Z(\beta )=\int [Dg]e^{-I^{E}[\beta,g]} \simeq e^{-I^{E}_{c}[\beta,g_{c}(\beta)]} \qquad \f{\delta I^{E}[\beta,g]}{\delta g}\Big |_{g=g_{c}(\beta)}=0,
\ee
where $I^{E}$ denotes the Euclidean action.
Then the entropy $S(\beta )$ of the
system is given by
\be
S(\beta)=\left( \beta  \f{\p}{\p \beta }-1\right) I^E[\beta,g_c(\beta)] .
\ee
Note that semiclassical metric  depends on $\beta$ because of  regularity.

There is an alternative way of computing $S(\beta)$,
\be
S(\beta)=\left( \alpha  \f{\p}{\p \alpha }-1\right) \Big|_{\alpha=1} I^{E}_{c}[\alpha \beta ,g_{c}(\beta) ] \label{eq:entd}.
\ee
This expression instructs to  first evaluate the Euclidean action for the fixed metric $g_c(\beta)$ but with varying time periodicity $\alpha \beta$.  Such geometries with $\alpha\neq 1$  have conical singularities, and their action can be calculated by the method reviewed in the previous section.
By substituting (\ref{eq:rs}) into (\ref{eq:entd}) one immediately obtains the Wald formula \cite{Solodukhin:2011gn,Fursaev:1995ef}.
\be
S=4\pi \int_{\Sigma}\sqrt{h}\f{\p\mathcal{L}}{\p R_{\mu \nu \alpha \beta}}n_{\mu \alpha} n_{\nu \beta}.\label{eq:wald}
\ee

Note that if we divide  the action into the regular part and the singular part, $ I^{E}_{c}[\alpha ,g_{c}(\beta) ] =I^{E}_{reg}+I^{E}_{sing}$,  the former
does not contribute to the entropy,
\be
\left( \alpha  \f{\p}{\p \alpha }-1\right)I^{E}_{reg}=0 \label{eq:van},
\ee
because $I^{E}_{reg}$ is proportional $\alpha$ as we saw in (\ref{eq:lim}).

\section{Entropy of black holes via conical singularities in the Chern-Simons formulation}\label{subsecCS}

In this section we adapt the discussion of the previous section to the Chern-Simons formulation.  We focus on BTZ black holes for our explicit computations, but then propose that the resulting entropy formula holds in general.   The validity of this proposal will be confirmed in the next section. 

\subsection{Entropy of the BTZ black hole}

It is well known that 3d Einstein gravity with negative cosmological constant can be formulated in terms of  Chern Simons theory with the gauge group $SL(2,R) \times SL(2,R)$,
\ba
I[A,\bar A]& =I_{CS}[A]-I_{CS}[\bar A], \\
 I_{CS}[A] &=\f{k}{4\pi} \int \mbox{Tr} \left(A  \wedge dA +\f{2}{3} A \wedge A \wedge A \right),
\end{align}
with $k=1/4G$.  For definiteness, in this section we consider the 2$\times 2$ matrix representation for SL(2,R) (although the final formulas will not end up depending on that choice) with generators obeying
\be
[L_i,L_j]=(i-j)L_{i+j}~,\quad \rm{Tr} (L_1 L_{-1}) = -1~,\quad \rm{Tr} (L_0 L_0)={1\over 2}~.
\ee
The connections are related to the vielbein $e$ and spin connection $\omega$ as
\be
A=\omega +e, \qquad \bar A=\omega -e,
\ee
and the metric is given by
\be
g_{\mu \nu} =2 \rm{Tr}[ e_{\mu} e_{\nu}].
\ee

The connections for the nonrotating BTZ black hole can be taken as
\begin{align}
A& = e^{\rho_{+}}\left( e^{r }L_{1}- e^{-r}L_{-1}\right) dx^{+}+ L_{0} dr  \\
  \bar A& =- e^{\rho_{+}}\left( e^{r}L_{-1}- e^{-r}L_{1}\right) dx^{-}- L_{0} dr \label{eq:btz},
\end{align}
where the horizon is at $r=0$  and $e^{\rho_{+}}=\s{\f{2\pi \lm}{k}}$, where $\lm$ is proportional to the mass.

The metric of the black hole is
\be
ds^2=4e^{2\rho_{+}}\left( -\sinh r ^2 dt^2+\cosh r^2 d\theta^2 \right) +dr^2.
\ee
 The value of the entropy of the black hole is obtained by the  Bekenstein-Hawking formula
\be
S=\f{A}{4G}=\f{4k\pi^2 }{\beta} \label{eq:bh},
\ee
where $\beta$ is the inverse Hawking temperature $\beta =\pi e^{-\rho_{+}}$.

We now show how to derive this by the conical singularity method.  If we keep the connections fixed but identify the time coordinate as $t \cong t+ i\alpha \beta$ then we introduce a conical singularity in the metric for $\alpha\neq 1$.  This can be seen at the level of the connections by evaluating the holonomies around the imaginary time circle,
\ba e^{\oint A} &= \left(\begin{array}{cc} \cos  \pi \alpha & -i e^{-r} \sin \pi \alpha \\
-i e^r \sin \pi \alpha &  \cos \pi \alpha  \end{array} \right)~, \\
 e^{\oint \overline{A}} &= \left(\begin{array}{cc} \cos  \pi \alpha & -i e^{r} \sin \pi \alpha \\
-i e^{-r} \sin \pi \alpha &  \cos \pi \alpha  \end{array} \right)~.
\end{align}
For a nonsingular metric we need the holonomies to be in the center of SL(2,R), which requires $\alpha$ to be an integer.

Now we would like to evaluate  $ I_{CS}[A] $ for this connection.  We proceed by regularizing the connection, evaluating its action, and then removing the regulator. 

It is convenient to use a rescaled  Euclidean time coordinate $T$, $t=i   \alpha \beta T$, so that the coordinate periodicity is fixed as  $T \cong  T +1$ .
In this coordinate the singular connection is written
\be
A=i \alpha \beta A_{t}dT+ A_{\theta}d\theta + L_{0}d\rho \qquad \bar{A}= i\alpha \beta \bar{A}_{t}dT+\bar{A}_{\theta}d\theta -L_{0}d\rho,
\ee
and the corresponding singular metric is
\ba
g_{TT}(S)&=-{1 \over 2} (\alpha \beta )^2\mbox{Tr} \left(A_{t}-\bar{A_{t}} \right)^2 \\ g_{\theta \theta}(S)&={1\over 2} \mbox{Tr} \left(A_{\theta}-\bar{A_{\theta}} \right)^2 .
\end{align}
There are various way to regularize the connection. For example, consider
\begin{align}
\tilde{A}&=\f{A_{t}}{u(r)}\left(i \beta\alpha dT -c d \theta \right)+\left( c A_{t}+A_{\theta} \right)d \theta + L_{0}d\rho  \nonumber \\
 \tilde{\bar{A}}&=\f{\bar{A}_{t}}{u(r)}\left( i\beta \alpha dT +c d \theta \right)+\left( -c \bar{A}_{t}+\bar{A}_{\theta} \right)d \theta -L_{0}d\rho, \label{eq:req}
\end{align}
where $c$ is some constant. The connections (\ref{eq:req}) are regular provided $ u(0)= \alpha$.
We also demand $u(\infty)= 1$
so that  we go back to the original  one at the boundary. $A_{\rho}$ and $\bar{A_{\rho}}$ are unchanged because we are working on the gauge where $A_{\rho}=L_{0},\bar{A}_{\rho}=-L_{0}$.

Below we fix the value of $c$ that appears in the connections so that the regularization is consistent with the regularization of the metric (\ref{eq:mmet}).
The metric components $g_{TT}(R)$ and $g_{\theta \theta}(R)$ of these regularized connection near the tip $r \sim 0$  look like

\ba
g_{TT}(R)&=-\f{1}{2}\beta^2\mbox{Tr} \left(A_{t}-\bar{A_{t}} \right)^2 \\
g_{\theta \theta}(R)&= \f{1}{2}\mbox{Tr}\left[c\left( A_{t}+\bar{A}_{t} \right)\left(1-\f{1}{\alpha}\right) +\left(A_{\theta}-\bar{A}_{\theta}\right)\right]^2 .
\end{align}
From these expressions one notices how the metric components change by the regularization. In particular,   in the  BTZ case,
\be
\delta g_{TT}= g_{TT}(R)-g_{TT}(S)=2(1-\alpha)g_{TT}(S), \qquad \delta g_{\theta \theta}=-2c(1-\alpha)g_{\theta \theta}(S).
\ee
We used the property  of the BTZ connection at $r=0$, namely, $A_{t}+\bar{A}_{t}=A_{\theta}-\bar{A}_{\theta}$.
As in the metric case (\ref{eq:mmet}), we demand that  the volume of the torus located at $r=r_{0},r_{0}<<1$  be held fixed. In this case, $c$ appearing
in the regularization (\ref{eq:req}) has to be 1.

Now that we have specified the regularization, we can compute the action $ I_{CS}[A] $ of the singular configuration via regularization. As we are only interested in terms
which are proportional to $(1-\alpha)$, we only have to calculate the $\int \mbox{Tr} A \wedge dA$ term, since the  $A^3$ term  is proportional to  $\alpha$  and vanishes in  (\ref{eq:entd})
\begin{align}
\f{k}{4\pi}\int \mbox{Tr}A \wedge dA& =\f{k}{4\pi}\int \mbox{Tr}\left[(A_{t}+A_{\theta})A_{t}\right] \wedge \left(-\f{u'(r)}{u(r)^2}dr \right)\wedge i\alpha \beta dT \wedge d \theta +I_{reg} \nonumber \\
&= -\f{i k\beta}{2} (1-\alpha)\mbox{Tr}\left[(A_{t}+A_{\theta})A_{t}\right] +I_{reg},
\end{align}
where $I_{reg}$ denotes the terms which are proportional to $\alpha$.
Similarly,
\be
\f{k}{4\pi}\int \mbox{Tr}\bar{A} \wedge d\bar{A}= -\f{i k\beta}{2} (1-\alpha)\mbox{Tr}\left[(-\bar{A_{t}}+\bar{A}_{\theta})\bar{A}_{t}\right] +\bar{I}_{reg}.
\ee
Since the Euclidean action is related to the Chern Simons actions via $i I_{E}=I_{CS}[A]-I_{CS}[\bar{A}]$, we derive the expression for the entropy of the BTZ black hole by using ($\ref{eq:entd})$,
\begin{align}
S&=\f{k\beta}{2} \mbox{Tr}\left[(A_{t}+A_{\theta})A_{t}\right]+\f{k\beta }{2} \mbox{Tr}\left[(\bar{A_{t}}-\bar{A}_{\theta})\bar{A}_{t}\right] \\ \nonumber
&=\f{4k\pi^2}{\beta}.
\end{align}
The result reproduces the Bekenstein Hawking formula (\ref{eq:bh}).

This result can be generalized to the rotating BTZ black hole.  In this case we have both an inverse temperature $\beta$ and the angular velocity of the horizon $\Omega$.  These can be combined to form
$\tau = {i\beta \over 2\pi}(1+\Omega)$ and $\taub = -{i\beta \over 2\pi}(1-\Omega)$.   For the Euclidean black hole, $\tau$ plays the role of the modular parameter of the boundary torus.   Repeating the above analysis for this case we find the result (see Appendix A)
\be
S=-2\pi i k \mbox{Tr}\left[A_{+}\left( \tau A_{+} - \bar{\tau}A_{-} \right)\right] -2\pi i k \mbox{Tr}\left[\bar{A}_{-}\left( \tau \bar{A}_{+} - \bar{\tau}\bar{A}_{-} \right)\right],
 \ee
which indeed yields the correct entropy of the rotating BTZ black hole.

Although this result was derived for the BTZ black hole, since the  formula does not make any specific reference to this solution we propose that it holds more generally.  It is not obvious that this is a correct assumption. In particular, in the above argument we didn't consider all possible regularizations of the singular connection, and the argument for setting the constant $c=1$ is not entirely compelling. Furthermore, the connection representing BTZ is not of the most general form.   Fortunately, we can check that the result is correct by verifying that it obeys the correct first law variation.  We carry this out in the next section.

\section{Derivation from the first law}\label{firstlaw}

In the preceding section we have motivated a simple expression for the entropy of a higher spin black hole.   In this section we wish to verify that the result is indeed correct, and can be applied to general  higher spin black holes.     Our main tool is the first law of thermodynamics: in the thermodynamic limit the entropy is defined to be the object whose variation satisfies the first law, and so if we can establish this property then we are done.

To keep the discussion as general as possible, we consider a theory with an infinite tower of higher spin charges, $(W_2, W_3, \ldots)$, where $W_s$ denotes a spin-s charge.  Here we focus on just the ``holomorphic" or "leftmoving" charges, but everything we say has an obvious parallel on the anti-holomorphic or rightmoving side.     Each conserved charge has a corresponding conjugate potential, and we denote these as $(\alpha_2, \alpha_3, \ldots)$.    We will interchangeably use a different notation for the spin-2 versions:  $W_2 \leftrightarrow \Lc$ and  $\alpha_2 \leftrightarrow \tau$, which are identified as the holomorphic stress tensor and modular parameter.

Following \cite{Gutperle:2011kf}  we think in terms of an underlying partition function of the form
\ba
Z &= {\rm Tr}  \left[ e^{4\pi^2 i \sum_{s=2}^\infty \alpha_s W_s} \right].
\end{align}
The right hand side has a precise meaning on the CFT side of the AdS/CFT correspondence, but here is just being used as a mnemonic for motivating the form of the first law.  Namely, we have
\ba
\delta S& = -4\pi^2 i \sum_{s=2}^\infty \alpha_s  \delta W_s ~.
\end{align}

Next, let us recall the general rules for constructing higher spin black holes, and identifying their charges and potentials.   We work in the context of hs$[\lambda] \times $ hs$[\lambda]$ Chern-Simons theory, and recall that upon setting $\lambda =\pm N$ this theory reduces to SL(N,R)$\times $SL(N,R) Chern-Simons theory.  In fact, it will become clear that our derivation will apply to any Lie algebra with an SL(2,R) subalgebra, which includes hs$[\lambda]$ as a special case.

The Lie algebra hs$[\lambda]$ has generators $V^s_m$, with $s=2, 3, \ldots$ and $m=-(s-1), \ldots s-1$.    An SL(2,R) subalgebra is furnished by $V^2_{\pm 1, 0}$.   The trace operation obeys
\ba
 {\rm Tr} (V^s_m V^t_n) \propto \delta_{s,t}\delta_{m,-n}
\end{align}
and in particular we write
\ba
 {\rm Tr} (V^s_{s-1}  V^s_{-(s-1)} )  =t_s~.
\end{align}
Another useful fact is that $V^2_1 V^s_{s-1}= V^s_{s-1} V^2_1 =V^{s+1}_s$.

As is standard we write the connection as
\ba
A= b^{-1} ab +b^{-1} db~,\quad b=e^{\rho V^2_0}
\end{align}
with
\ba
a = a_z dz + a_{\zb} d\zb~.
\end{align}
Note that we are working in Euclidean signature.  The component $a_z$ is taken to be in highest weight gauge \cite{Campoleoni:2010zq}
\ba
a_z = V^2_1 + \sum_{s=2}^\infty c_s W_s V^s_{-(s-1)}~.
\end{align}
The constants $c_s$ are fixed by demanding that the charges  $W_s$ obey the algebra of W$_\infty[\lambda]$. In particular, $c_2$ is fixed to be
\ba
c_2 = {2\pi \over t_2 k}~.
\end{align}

Next we need to specify $a_{\zb}$.  To define a flat connection it has to commute with $a_z$, which can be satisfied by taking $a_{\zb}$ to depend on powers of $a_z$,
\ba
a_{\zb} = \sum_{s=2}^{\infty} f_{s+1} (a_z)^s\Big|_{\rm traceless}
\end{align}
where $f_s$ are coefficients.
The $s=1$ term is absent, since it can be removed by redefining the coordinates $(z,\zb)$. Noting the property $(V^2_1)^s = V^{s+1}_s$ we can write
\ba
a_{\zb} = \sum_{s=2}^\infty f_{s+1} ( V^{s+1}_s +\ldots )
\end{align}
where $\ldots$  denote generators with small value of the lower mode index.   Restoring the $\rho$ dependence, the leading terms displayed above give the leading large $\rho$ behavior.

The coefficients $f_s$ are fixed by the holonomy conditions. The Euclidean black hole has coordinates identified as $(z,\zb) \cong (z+2\pi, \zb +2\pi) \cong (z+2\pi \tau, \zb+2\pi \taub)$.  Assuming constant $a$, the holonomy around the $\tau$ cycle is
\ba
H = e^\omega~,\quad \omega =2\pi(\tau a_z + \taub a_{\zb})~.
\end{align}
A smooth solution is obtained provided $H$ lies in the center of the gauge group, which requires that $\omega$ has certain fixed eigenvalues.  We can impose these conditions by requiring the ${\rm Tr}(\omega^n)$, $n=2, 3, \ldots$,  take fixed values; for instance, for black holes smoothly connected to BTZ, we demand that these traces coincide with their BTZ values.   These equations are in one-to-one correspondence to the free parameters $f_s$, and can be used to fix their values.

The constants $f_s$ are proportional to the potentials $\alpha_s$ appearing in the first law.  This was originally shown by a Ward identity analysis \cite{Gutperle:2011kf}.   As will become clear momentarily, the relation is
\ba
f_s =- {2\pi \over k c_s t_s} {\alpha_s \over \taub}
\end{align}
so that we have
\ba
a_{\zb} = -{2\pi \over k} \sum_{s=2}^\infty {\alpha_s \over c_s t_s \taub}  ( V^{s+1}_s +\ldots )~.
\end{align}
The holonomy relations can now be used to express the potentials $\alpha_s$ in terms of the charges $W_s$, or vice versa.

Written in terms of the holonomy, the candidate entropy formula is
\ba S = -ik  {\rm Tr}(a_z \omega)
\end{align}
together with its anti-holomorphic partner. We now verify that this obeys the correct first law of thermodynamics.  We have
\ba \delta S = -ik {\rm Tr} ( \delta a_z \omega) - ik {\rm Tr}( a_z  \delta\omega)~.
\label{varS}
\end{align}
However, it's easy to see that the second contribution vanishes.  The condition that the traces of $\omega$ take fixed values implies that $\delta \omega = [\omega, X]$ for some $X$.  Using this, along with $[a_z,\omega]=0$, which follows from the fact that $\omega$ is built out of powers of $a_z$, we readily verify ${\rm Tr }(a_z \delta \omega)=0$.  The variation of $a_z$ is
\ba
\delta a_z = \sum_{s=2}^\infty c_s \delta W_s V^s_{-(s-1)}~.
\end{align}
Inserting this and taking the trace yields
\ba
\delta S& = -4\pi^2 i \sum_{s=2}^\infty \alpha_s  \delta W_s ~.
\end{align}
as desired.

Without doing any computations, we can establish that our entropy formula will agree with the results obtained in  \cite{Gutperle:2011kf,Castro:2011fm,Tan:2011tj,David:2012iu,Kraus:2011ds}.   This is because those computations were based on solving the holonomy conditions and then integrating the first law variation.   Here we have shown that if the holonomy conditions are imposed then our entropy formula obeys the correct first law variation.  Therefore, it must agree with previous results.

Let us make some further comments.   In the original work \cite{Gutperle:2011kf} it appeared somewhat miraculous that that the first law variation could be consistently integrated; this required verifying the integrability constraints, which turned out to follow rather non-transparently from the holonomy condition.  Our discussion here removes the mystery surrounding this procedure.  In particular in (\ref{varS}) we see very explicitly that if the holonomy is not kept fixed then $\delta S$ will acquire an additional unwanted term.    This makes it  clear that one should fix the holonomy in order to obtain the desired first law. Note also that any fixed holonomy will do, in terms of satisfying the first law.

 Another notable point is that our  derivation extends essentially automatically to to an arbitrary gauge group containing an SL(2,R) subgroup.  Simply decompose the the generators into irreducible representations of SL(2,R), and denote the generators in a given representation as ${V_{(i)}}^s_m$, $m=-(s-1), \ldots s-1$,  where the $i$ label takes into account the multiplicity of a given representation.   Note though that in our discussion we took the index $s$ to obey $s>2$, which leaves out the singlet; including the $s=1$ case is straightforward.   For some gauge groups, such as SL(N,R) with $N>2$, there are multiple inequivalent choices of SL(2,R) subgroups, and these lead to the existence of black holes with different asymptotics \cite{Ammon:2011nk}.  From the CFT point of view, these correspond to thermal states in CFT with different W-algebra symmetries.  Since nowhere in our computation did we assume a particular SL(2,R) subgroup, it should be clear that our entropy formula applies to all such cases.

\section{Conclusion}

In this paper we obtained a formula that computes the entropy of a
general stationary higher spin black hole in 2+1 dimensions. Although the notion of an event horizon
is somewhat vague for  black holes in higher spin theories, we do have a definite notion of a conical
singularity in terms of holonomy.  A Euclidean  black hole with the period of
the timelike cycle different from the inverse of the Hawking temperature is an example of a
configuration with a conical singularity.  Then the field strength of the configuration is delta function divergent at the singularity. It was recognized \cite{Banados:1993qp,Iyer:1995kg,Solodukhin:2011gn,Fursaev:1995ef}  that only
the contribution from the conical singularity to the action is necessary to reproduce  the entropy of the black hole.

With this point of view, in this paper we developed a method to calculate the contribution of the conical singularity to the Chern Simons action, by carefully
regularizing  the connection.  We used this result to compute the black hole entropy. 
An advantage of this method is that since only terms with radial derivatives  in the action have a
 chance to produce a delta function like divergence, it is sufficient to evaluate the $ \int A \wedge dA $ term in the Chern Simons action. We don't need to
care about other terms, such as boundary terms,  which are required when evaluating the total free energy of the black hole. More precisely, since these terms are
all proportional to $\alpha$ (deficit parameter), they vanish in (\ref{eq:entd}).

This statement is equally true in the metric formulation of Einstein gravity. When we evaluate the free energy of  the asymptotically flat Schwarzchild black hole,
 the bulk Einstein-Hilbert action vanishes and we have to take into account the  Gibbons-Hawking boundary term. On the other hand,
when we compute the entropy of the black hole by the conical singularity method, we only have to evaluate the contribution
of the conical singularity at the tip to the Einstein-Hilbert action.  Since the entropy only depends on the local geometry of the horizon, it is efficient to use a computational scheme that makes this fact manifest.

It is useful to compare the general status of our entropy formula with that of the area law or Wald formula.    A nice property of the latter is that they can be evaluated on any cross section of the horizon.   Therefore, they make sense even when applied to non-stationary black holes, such as those that are absorbing infalling matter, although the physically correct formula for dynamical black hole entropy might need to include additional terms in order to guarantee its monotonicity in time.   On the other hand, our formula depends explicitly on the parameter $\tau$, which depends on the black hole temperature.    Since the notion of temperature only makes sense in thermodynamic equilibrium, we don't expect to be able to apply our result to out of equilibrium black holes.    An outstanding challenge is therefore to find an entropy formula which does make sense out of equilibrium.

Another feature that could be improved is that our formula only applies to black holes and not thermal AdS.   If one blindly plugs in the gauge connection for thermal AdS into our entropy formula one finds a nonzero result, whereas the correct answer is of course zero.   The area law or Wald entropy automatically assign zero entropy to thermal AdS, simply because there is no event horizon.  The reason why this is inconvenient is the following.   Given the entropy, the partition function is obtained via Legendre transformation as $\ln Z = S + 4\pi^2i  \sum_s \alpha_s W_s$.   For BTZ black holes a very useful observation is that Euclidean BTZ and thermal AdS are related by a coordinate transformation that acts as a modular transformation on the boundary.   The partition function is obtained from the Euclidean action and so is invariant under coordinate transformations, and this makes modular invariance of the partition function manifest.   Among other things, this provides a very convenient way of computing the partition function and entropy of BTZ.  This strategy fails when applied to our formalism, because our entropy formula, and hence the partition function derived from it, only applies to black holes and not thermal AdS.  Thus we cannot use it to establish the modular properties of the partition function, which would be very useful in order to make direct contact with the CFT.

\section*{Acknowledgements}
We would like to thank T. Azeyanagi, J. de Boer, D. Fursaev, S. Hellerman, E. Perlmutter for discussions, and T.Takayanagi for reading the manuscript carefully, frequent discussions, encouragement and support.  P.K. is supported in part by NSF grant PHY-07-57702.
T.U. is supported by World Premier International Research Center Initiative (WPI
Initiative), MEXT, Japan, and by JSPS Research Fellowships for Young Scientists.

\section*{Appendix A. Entropy formula for general stationary black holes}
In this appendix  we  derive an entropy formula for general stationary higher
spin black holes. An example of a stationary black hole is a rotating BTZ black hole. The connection
of the black hole is given by
\begin{align}
A&=\left(e^{\rho}L_{1}-e^{-\rho}\f{2\pi \mathcal{L} }{k}L_{-1} \right)dx^{+} +L_{0} d\rho \\ \nonumber
 \bar{A}&=-\left(e^{\rho}L_{-1}-e^{-\rho}\f{2\pi \bar{\mathcal{L}} }{k}L_{1} \right)dx^{+} -L_{0} d\rho
\end{align}
with $\mathcal{L} \neq \bar{\mathcal{L}}$.  The event horizon is located at
\be
e^{2\rho_{+}}=\f{2\pi}{k}\sqrt{\lm \bar{\mathcal{L}}}.
\ee
The entropy of the black hole is given by
\be
S=\f{A}{4G}=2\pi \left(\s{2\pi \lm k}+\s{2\pi \bar{\lm}k} \right).
\label{srot}
\ee
In the corresponding Euclidean configuration, the modular parameter $\tau$
of the boundary torus contains non vanishing real part,
\be
\tau=\f{i\beta}{2\pi}\left(1+\Omega \right)=\f{ik}{2} \f{1}{\s{2\pi k \lm}},
\label{tauval}
\ee
where  $\Omega$ is the complex
angular velocity.
Since lines $\Theta =\theta -\Omega t =const$ are contractible cycles, it is convenient
to write the connection as
\begin{align}
 A_{t}dt+A_{\theta}d \theta& =i\beta \left(A_{t}+\Omega A_{\theta} \right)dT +A_{\theta}d \Theta \nonumber \\
&=2\pi  \left (\tau A_{+}-\bar{\tau} A_{-}\right) dT +A_{\theta}d \Theta.
\end{align}
We introduced a rescaled Euclidean time coordinate $T$ which satisfies $t=i \beta T$ and $T \cong T+1$.

The holonomy around  the contractible cycle
is derived by integrating the connection along the line $\Theta =0 $.  When (\ref{tauval}) is satisfied the holonomy lies in the center of the gauge group.
Suppose we replace $\beta \rightarrow \alpha \beta$ appearing in the connection and vary $\alpha$ away from 1 
 while the relation $ T \sim T+1$ is kept fixed. Then the connection develops a conical singularity and
the holonomy becomes nontrivial. To evaluate the action, we have to regularize the connections.
It turns out that the connections
\begin{align}
\tilde{A}& =\f{1}{u(r)}\left(2\pi \left( \tau A_{+} - \bar{\tau}A_{-} \right)dT -A_{t}d\Theta \right)+\left(A_{t}+A_{\theta} \right)d \Theta \nonumber \\
\tilde{\bar{A}}& =\f{1}{u(r)}\left(2\pi \left( \tau \bar{A}_{+} - \bar{\tau}\bar{A}_{-} \right)dT +\bar{A}_{t}d\Theta \right)+\left(-\bar{A}_{t}+\bar{A}_{\theta} \right)d \Theta
\end{align}
are the right regularization because they do not change the volume of the  torus located at $\rho, \rho-\rho_{+}<<1$ for the rotating BTZ black hole.
By evaluating the $\int A \wedge dA$ term and taking the derivative in terms of $\alpha$, we get  the final expression,
\be
S=-2\pi ik  {\rm Tr} \left[A_{+}\left( \tau A_{+} - \bar{\tau}A_{-} \right)\right] -2\pi i k {\rm Tr}\left[ \bar{A}_{-}\left( \tau \bar{A}_{+} - \bar{\tau}\bar{A}_{-} \right)\right].
 \ee
It is straightforward to verify that this yields agreement with (\ref{srot}).

\end{document}